\documentclass[12pt]{article}

\newcommand{\EQ}{\begin{equation}}
\newcommand{\EN}{\end{equation}}

\newcommand{\ket}[1]{\left|#1\right\rangle}      

\newcommand{\bear}{\begin{eqnarray}}
\newcommand{\ear}{\end{eqnarray}}
\newcommand{\bt} { \begin{tabular} }
\newcommand{\et}{ \end{tabular} }
\newcommand{\bc} { \begin{center} }
\newcommand{\ec}{ \end{center} }

\newcommand{\btb} { \begin{table} }
\newcommand{\etb}{ \end{table} }

\begin{document}

\topmargin 0pt
\oddsidemargin 5mm
\newcommand{\NP}[1]{Nucl.\ Phys.\ {\bf #1}}
\newcommand{\PL}[1]{Phys.\ Lett.\ {\bf #1}}
\newcommand{\NC}[1]{Nuovo Cimento {\bf #1}}
\newcommand{\CMP}[1]{Comm.\ Math.\ Phys.\ {\bf #1}}
\newcommand{\PR}[1]{Phys.\ Rev.\ {\bf #1}}
\newcommand{\PRL}[1]{Phys.\ Rev.\ Lett.\ {\bf #1}}
\newcommand{\MPL}[1]{Mod.\ Phys.\ Lett.\ {\bf #1}}
\newcommand{\JETP}[1]{Sov.\ Phys.\ JETP {\bf #1}}
\newcommand{\TMP}[1]{Teor.\ Mat.\ Fiz.\ {\bf #1}}

\renewcommand{\thefootnote}{\fnsymbol{footnote}}

\newpage
\setcounter{page}{0}
\begin{titlepage}
\begin{flushright}
UFSCARF-TH-09-10
\end{flushright}
\vspace{0.5cm}
\begin{center}
{\large Exactly solvable models of interacting spin-$s$ particles in one-dimension}\\
\vspace{1cm}
{\large C.S. Melo and M.J. Martins} \\
\vspace{1cm}
{\em Universidade Federal de S\~ao Carlos\\
Departamento de F\'{\i}sica \\
C.P. 676, 13565-905~~S\~ao Carlos(SP), Brasil}\\
\end{center}
\vspace{0.5cm}

\begin{abstract}
We consider the exact solution of a many-body problem of spin-$s$ particles interacting through
an arbitrary U(1) invariant factorizable $S$-matrix.
The solution is based on a unified formulation of the quantum inverse
scattering method for an arbitrary $(2s+1)$-dimensional monodromy matrix.
The respective eigenstates are shown to be given in terms of $2s$ creation fields by a general new recurrence relation.
This allows us to derive the spectrum and the respective Bethe ansatz equations.
\end{abstract}

\vspace{.15cm}
\centerline{PACS numbers:  05.30-d, 04.20Jb, 02.30.Ik}
\vspace{.1cm}
\vspace{.15cm}
\centerline{October 2006}

\end{titlepage}


\pagestyle{empty}

\newpage

\pagestyle{plain}
\pagenumbering{arabic}

\renewcommand{\thefootnote}{\arabic{footnote}}

The existence of exactly solved models have been
playing a fundamental role in our understanding of interacting
one-dimensional many-particle systems in distinct  areas of
theoretical physics \cite{BA,GA,DC,KON}. An important class of
models are those constituted by $n_{s}$ species of particles of
identical mass that interact only through a sequence of two-body
scattering processes. The two-body collision amplitudes
$S(\mu_1,\mu_2)_{b_1,b_2}^{a_1,a_2}$, where $a_j,b_j=1,\dots,n_s$
represent internal quantum numbers, are responsible for the
redistribution of momenta among the different types of particles
\cite{ZA}. In general, these theories lead us to deal with
non-diagonal scattering matrices defined by,
\begin{equation}
\hat{S}_{12}(\mu_1,\mu_2)=\sum_{a_1,a_2,b_1,b_2}^{n_s}S(\mu_{1},\mu_{2})_{b_{1},b_{2}}^{a_{1},a_{2}}
e_{b_{1},a_{1}} \otimes e_{b_{2},a_{2}}, \label{SM}
\end{equation}
where $e_{a,b}$ denotes $n_s \times n_s$ 
matrices 
having only
non-null element with value 1 at row $a$ and column $b$.

The rapidity $\mu_j$ parameterizes the one-particle momenta
$p(\mu_j)$ and energy $\epsilon(\mu_j)$ associated to the $jth$
particle. We stress here that we will consider the most general
situation in which the $S$-matrix (\ref{SM}) depends on both
values of two independent $\mu_1$ and $\mu_2$ scattering
parameters. The scattering from the initial state $(a_{1},a_{2})$
to the final state $(b_{1},b_{2})$ may exchange quantum numbers
and momenta but do not alter the values of the latter parameters.
This type of factorizable theory has been first emerged in the
study of the eigenspectrum problem associated to particles
interacting through delta-function potential \cite{LI,MA,YY}. It
was then discovered that in order to assure exact solubility
the $S$-matrix (\ref{SM}) needed to
satisfy a necessary condition called Yang-Baxter equation
\cite{MA,YY},
\begin{equation}
S(\mu_{1},\mu_{2})_{a_{1},a_{2}}^{\gamma_{1},\gamma_{2}}
S(\mu_{1},\mu_{3})_{\gamma_{1},a_{3}}^{b_{1},\gamma_{3}}
S(\mu_{2},\mu_{3})_{\gamma_{2},\gamma_{3}}^{b_{2},b_{3}} =
S(\mu_{2},\mu_{3})_{a_{2},a_{3}}^{\gamma_{2},\gamma_{3}}
S(\mu_{1},\mu_{3})_{a_{1},\gamma_{3}}^{\gamma_{1},b_{3}}
S(\mu_{1},\mu_{2})_{\gamma_{1},\gamma_{2}}^{b_{1},b_{2}}.
\label{p2}
\end{equation}
where sum of repeated indices is
assumed. Here we shall consider solutions of (\ref{p2}) that are
almost unitary, namely
\begin{equation}
S(\mu_1,\mu_2)_{a,b}^{c,d}
S(\mu_2,\mu_1)_{d,c}^{\alpha,\gamma}=\rho(\mu_1,\mu_2)
\delta_{a,\gamma} \delta_{b,\alpha},
\label{p3}
\end{equation}
for some arbitrary function $\rho(\mu_1,\mu_2)$.

Assuming the existence of such integrable systems, the next step
is to quantize them to determine the respective eigenspectrum. As
first argued by Sutherland \cite{SU} this can be accomplished
within the framework of an asymptotic Bethe ansatz approach. The
basic idea is that in the regions of the configuration space where
the distance between the particles is much larger than the
interaction range, the wave-function can be considered as that of a
superposition of plane waves. It can be constructed similarly as
for the problem of particles with delta-interaction \cite{YY}, by
replacing the scattering matrix of the delta-potential by a
general factorizable $S$-matrix (\ref{SM}). If we consider
periodic boundary conditions on a ring of size $L$, the asymptotic
one-particle momenta $p(\mu_j)$ is required to satisfy the
following eigenvalue equation,
\begin{equation}
\mbox{e}^{ i p(\mu_{j}) L
}=\frac{\Lambda(\lambda=\mu_{j},\overrightarrow{\mu})}{[\rho(\mu_{j},\mu_{j})]^{1/2}},~j=1,\dots,N,
\label{BAE}
\end{equation}
where $N$ is the number of particles in the system and
$\overrightarrow{\mu}$ denotes the set $\lbrace
\mu_1,\mu_2,\dots,\mu_N \rbrace $ of corresponding rapidities.
The energy $E$ of
the system is given by adding the one-particle
energy expressions, namely
\begin{equation}
E= \sum_{j=1}^{N} \epsilon(\mu_j).
\end{equation}

The function $\Lambda(\lambda,\overrightarrow{\mu})$ represents
the eigenvalues of an auxiliary $[n_{s}]^{N} \times [n_{s}]^{N}$
operator $T(\lambda,\overrightarrow{\mu})$ usually called transfer
matrix. This auxiliary eigenvalue problem can be defined by,
\begin{equation}
T(\lambda,\overrightarrow{\mu}) \ket{\psi}=\mathrm{Tr}_{\cal{A}}
\left [ \cal{T}_{\cal{A}} (\lambda,\overrightarrow{\mu}) \right ]
\ket{\psi} = \Lambda(\lambda,\overrightarrow{\mu}) \ket{\psi},
\label{p4}
\end{equation}
where the trace is taken over an auxiliary $n_s$-dimensional space
${\cal{A}} \equiv C^{n_s}$. Furthermore, the monodromy operator
$\cal{T}_{\cal{A}} (\lambda,\overrightarrow{\mu}) $ is related to
the $S$-matrix elements by the following ordered product
\cite{QS,KO},
\begin{equation}
{\cal{T}}_{\cal{A}} (\lambda,\overrightarrow{\mu})=
{\hat{S}}_{{\cal{A}} N}(\lambda,\mu_N) {\hat{S}}_{{\cal{A}}
N-1}(\lambda,\mu_{N-1}) \dots {\hat{S}}_{{\cal{A}}
1}(\lambda,\mu_{1}). \label{p5}
\end{equation}

In order to make further progress it becomes crucial the exact
solution of the eigenvalue problem (\ref{p4},\ref{p5}) for
arbitrary $n_{s}$ and two-body scattering amplitudes. This is
indeed  a tantalizing open problem, especially when
a particular
form of
$S(\mu_1,\mu_2)_{a,b}^{c,d} $ is not specified. Indeed,
most of the results concentrate on
specific $S$-matrices such as those  related to the
the six-vertex model \cite{LIE}, to 
its higher spin descendents \cite{BAB} and those based on higher
rank Lie algebras \cite{SU1,DE,PB,WI,JA}. In the latter case, 
some of the findings \cite{WI,JA} are still in
the form of conjectures for the transfer matrix eigenvalues and also
the number $n_{s}$ actually encodes a variety of
distinct conserved quantum numbers such as spin,
color, flavor, etc. Consequently, the number of null
scattering coefficients grows rapidly with $n_{s}$ due to the many
possible underlying $U(1)$ symmetries.

In this work, however, we shall establish the essential tools to
solve the eigenvalue problem (\ref{p4},\ref{p5}) when only a
unique $U(1)$ symmetry is present for arbitrary $n_{s}$. This is
the minimal continuous invariance one could request and our
results will be valid for arbitrary factorizable
$S$-matrices  satisfying such
symmetry condition and the unitarity relation (\ref{p3}). More
precisely, we are considering integrable
models whose $S$-matrices fulfill the property,
\begin{equation}
\left [ \hat{S}_{12}(\mu_1,\mu_2), S_1^{z}+S_2^{z} \right ]=0,
\label{sym}
\end{equation}
where $S_{j}^{z}$ is the azimuthal component of spin-s operator
associated to
the $jth$ particle such that $s=(n_s-1)/2$. Note that relation
(\ref{sym}) means $S(\lambda,\mu)_{b_{1},b_{2}}^{a_{1},a_{2}}=0$
unless the ice rule $a_{1}+a_{2}=b_{1}+b_{2}$ is satisfied which
leads us to a total number of $n_{s}(2n_{s}^{2}+1)/3$ non-null
amplitudes.

The direct connection between the number of species and the spin
of the particles makes these integrable models physically
relevant. We shall tackle this problem by means of the quantum
inverse scattering method \cite{QS,KO}. We remark that, in general, there is no
known recipe to perform this task, therefore our results can be
considered as new developments in this framework.

The most important quantity in this method are the monodromy
matrix elements on the auxiliary space which here will be shortly
denoted by ${\cal T}(\lambda)_{p,q}$, $p,q=1,\dots,n_s$.  These
are operators on the quantum space $ \prod_{j=1}^{N}\otimes
C_{j}^{n_s}$ such that the diagonal entries define the transfer
matrix eigenvalue problem (\ref{p4}) while the off-diagonal ones
will play the role of creation and annihilation fields. The
commutation relations between such matrix elements are given with
the help of the corresponding Yang-Baxter algebra,
\begin{equation}
S(\lambda,\mu)_{a,b}^{\alpha,\gamma} {\cal
T}_{\alpha,p}(\lambda){\cal T}_{\gamma,q}(\mu)={\cal T}_{b,c}(\mu)
{\cal T}_{a,d}(\lambda) S(\lambda,\mu)_{d,c}^{p,q}. \label{p6}
\end{equation}

In order that the model be soluble by means of an algebraic Bethe
ansatz, it is fundamental the existence a vacuum state $\ket{0}$
such that the monodromy operator (\ref{p5}) acts on it as a
triangular matrix on the auxiliary space for arbitrary $\lambda$.
Thanks to the underlying $U(1)$ symmetry, it is possible to build
up this state by the tensor product of local vectors,
\begin{equation}
\ket{0}=\prod_{j=1}^{N} \otimes \ket{s}_{j},~S_{j}^{z}
\ket{s}_{j}=s \ket{s}_{j} \label{vaccum},
\end{equation}
where $\ket{s}_{j}$ denotes the $jth$ spin-s highest state vector.

It turns out that this is one possible eigenvector on which the
matrix elements of ${\cal T}_{p,q}(\lambda)$ operate as follows

\begin{eqnarray}
{\cal T}_{p,q}(\lambda)\ket{0}= \left \{ \begin{array}{ll}
               \displaystyle{
\prod_{j=1}^{N} S(\lambda,\mu_{j})_{p,1}^{p,1} \ket{0}}
& \mbox{ for}~~ p=q \\
               \displaystyle{0}
& \mbox{ for}~~ p < q \\
               \displaystyle{\ket{pq}}
& \mbox{ for}~~ p < q

                \end{array} \right.
\label{vac}
\end{eqnarray}
where $\ket{pq}$ are non-null vectors representing excitations
over the vacuum $\ket{0}$.

From the above expressions we see that the fields ${\cal
T}_{p,q}(\lambda)$ for $p>q$ acts as creation operators with
respect to the reference state $\ket{0}$. Of particular importance
are the operators ${\cal T}_{1,q}(\lambda)$ ($q \ge 2$) which
satisfy the following property,
\begin{equation}
\left[{\cal T}_{1,q}(\lambda),\sum_{j=1}^{N} S_{j}^{z}
\right]=(q-1) {\cal T}_{1,q}(\lambda). \label{p10}
\end{equation}

A direct consequence of (\ref{p10}) is that the fields ${\cal
T}_{1,q}(\lambda)$ can be interpreted as raising operators
associated to excitations over the ferromagnetic vacuum $\ket{0}$
with spin component $s-q+1$. It is therefore plausible to suppose
that the eigenvectors
$\ket{\psi}=\Phi_{m}(\lambda_{1},\dots,\lambda_{m})\ket{0}$ of
(\ref{p4}) should be constructed similarly as a Fock space having
$2s$ possible distinct creation fields. Considering that both the
total number of particles and spin are conserved quantities one
expects that the eigenvectors structure should be as follows. The
field ${\cal T}_{1,2}(\lambda_{1})$ represents the one-particle
state, the linear combination ${\cal T}_{1,2}(\lambda_{1}) {\cal
T}_{1,2}(\lambda_{2})+\psi_{1}(\lambda_{1},\lambda_{2}) {\cal
T}_{1,3}(\lambda_{1}) {\cal T}_{1,1}(\lambda_{2})$, for some
function $\psi_{1}(\lambda_{1},\lambda_{2})$, the two-particle
states and so forth. It turns out that the form of such linear
combinations can be inferred on the basis of the commutation rules
between the fields ${\cal T}_{1,q}(\lambda)$ derived from the
quadratic algebra (\ref{p6}). In addition to that, we emphasize
that much of the simplifications needed to construct suitable
eigenstates are carried out only on basis of an extensive use of the
Yang-Baxter (\ref{p2}) and the unitarity (\ref{p3}) constraints between the scattering
amplitudes. Omitting here the technicalities of these computations
\cite{CM} we find that the multi-particle states obey a $2s$-order
recursion relation given by,
\begin{eqnarray}
\Phi_{m}( \lambda_{1},\dots,\lambda_{m}) &=&\sum_{k=1}^{min\lbrace
2s,m \rbrace} \sum_{2 \le j_{2}<\dots<j_{k} \le m}
\psi_{\frac{k}{2}}(\lambda_1,\lambda_{j_2},\dots,\lambda_{j_k})
{\cal T}_{1,1+k}(\lambda_{1})
\Phi_{m-k}(\lambda_2,\dots,\lambda_{j_2-1},\lambda_{j_2+1},
\nonumber \\
&& \dots,
\lambda_{j_3-1},\lambda_{j_3+1},
\dots, \lambda_{j_k-1},\lambda_{j_k+1}, \dots,\lambda_m) \times
\prod_{d=2}^{k} {\cal T}_{1,1}(\lambda_{j_{d}}) \prod_{l=2}^{k}
\nonumber \\
&& \times \prod_{\stackrel{t_{1}=2}{t_{1}\neq j_{2},\dots, j_{k}}}^{m}
\frac{S(\lambda_{t_{1}},\lambda_{j_{l}})_{1,1}^{1,1}}{S(\lambda_{t_{1}},\lambda_{j_{l}})_{2,1}^{2,1}}
\prod_{\stackrel{t_{2}=2}{t_{2}\neq j_{2},\dots, j_{k}}}^{j_{l}}
\Theta (\lambda_{t_{2}},\lambda_{j_{l}}), \label{p11}
\label{state}
\end{eqnarray}
where we identify $\Phi_{0}\equiv 1$ and
function $\Theta(\lambda,\mu)$ is,
\begin{equation}
\Theta(\lambda,\mu)=\frac{S(\lambda,\mu)_{3,1}^{3,1}
S(\lambda,\mu)_{2,2}^{2,2}-S(\lambda,\mu)_{3,1}^{2,2}
S(\lambda,\mu)_{2,2}^{3,1}}{S(\lambda,\mu)_{1,1}^{1,1}
S(\lambda,\mu)_{3,1}^{3,1}}.
\end{equation}

In the course of our analysis we also have made the natural
hypothesis that the fields states (\ref{state}) are symmetric
functions in all $\lbrace \lambda_{j} \rbrace$ variables. This
exchange property between the variable $\lambda_1$ and $\lambda_2$ can be
used to determine
$\psi_{\frac{k}{2}}(\lambda_1,\lambda_{j_2},\dots,\lambda_{j_k})$
in terms of the scattering amplitudes
recursively.
The simplest case being $s=1/2$ where by 
fixing $\psi_{\frac{1}{2}}(\lambda)=1$ we find that 
$\psi_{1}(\lambda,\mu)$ is,
\begin{equation}
\psi_{1}(\lambda,\mu)=-
\frac{S(\lambda,\mu)_{3,1}^{2,2}}{S(\lambda,\mu)_{3,1}^{3,1}}.
\end{equation}

Similar task for higher spin is in general more
involved but it can be done for any specific value of the spin \cite{CM}.
As an extra example we present below the explicit expression for $s=3/2$,
\begin{eqnarray}
\psi_{\frac{3}{2}}(\lambda,\mu,\tau) & = & \frac{\Theta(\lambda,\mu)
\psi_1(\lambda,\tau) S(\lambda,\mu)_{1,1}^{1,1}
S(\lambda,\mu)_{4,1}^{3,2}} {S(\lambda,\mu)_{3,2}^{4,1}
S(\lambda,\mu)_{4,1}^{3,2} -S(\lambda,\mu)_{3,2}^{3,2}
S(\lambda,\mu)_{4,1}^{4,1}}
\nonumber \\
&& + \psi_{1}(\mu,\tau) \frac{S(\lambda,\mu)_{3,2}^{2,3}
S(\lambda,\mu)_{4,1}^{3,2}- S(\lambda,\mu)_{3,2}^{3,2}
S(\lambda,\mu)_{4,1}^{2,3}} {S(\lambda,\mu)_{3,2}^{4,1}
S(\lambda,\mu)_{4,1}^{3,2}- S(\lambda,\mu)_{3,2}^{3,2}
S(\lambda,\mu)_{4,1}^{4,1}}.
\end{eqnarray}

For $s=1/2$ our result (\ref{p11}) recovers
the known algebraic Bethe states associated
to the six-vertex model \cite{QS,KO} while for $s=1$ we have an
extension of the Bethe states for nineteen-vertex  models
\cite{TA}. In fact, apart from factorizability and unitarity,
no other
assumptions on the $S$-matrices elements
entering the eigenstates
(\ref{p11})
have been made. In addition,
the generality of our recursive manner
to generate the eigenstates
for  arbitrary spin-s
is, as far as we know,
a new progress
in the quantum inverse scattering approach . It follows from
commutation relation (\ref{p10})
and our construction (\ref{p11}) 
the property,
\begin{equation}
\sum_{j=1}^{N}S_{j}^{z} \Phi_{m}( \lambda_{1},\dots,
\lambda_{m})\ket{0}=(sN-m) \Phi_{m}( \lambda_{1},\dots,
\lambda_{m})\ket{0},
\end{equation}
which corroborates the physical interpretation of
$\Phi_{m}(\lambda_{1},\dots, \lambda_{m})\ket{0}$ being
multi-particle states over $\ket{0}$.

We now turn our attention to determination of the eigenvalues
$\Lambda(\lambda,\overrightarrow{\mu})$. From (\ref{p4}) one has
to carry on the diagonal fields ${\cal T}_{p,p}(\lambda)$ over the
creation fields ${\cal T}_{1,q}(\mu)$ ($q \ge 2$) that built the
multi-particle states (\ref{p11}). This task is made by
recasting the Yang-Baxter algebra (\ref{p6}) in the form of commutation rules
between these fields. In general, convenient commutation rules
do not follow immediately from (\ref{p6}) and a two-step
procedure is needed. To give an example of our approach let us
denote by $[l;k]$ the $lth$ row and the $kth$ column of
(\ref{p6}). The appropriate commutation relations between the
fields ${\cal T}_{1,2}(\lambda)$ and ${\cal T}_{p,p}(\lambda)$ for
$1<p<n_{s}$ are obtained by using the combination $[l;(l-1)*n_{s}+2]
S_{l+1,1}^{l+1,1}(\lambda,\mu) - [l;l*n_{s}+1]
S_{l+1,1}^{l,2}(\lambda,\mu)$ of the entries of (\ref{p6}). The
basic idea is to keep the diagonal operator ${\cal
T}_{p,p}(\lambda)$ always in the right-hand side position in the
commutation rules and such procedure can be implemented for all
creation fields ${\cal{T}}_{1,q}(\mu)$. The eigenvalues are easily
collected by keeping only the first terms of the commutation rules
among ${\cal{T}}_{1,q}(\mu)$ and ${\cal{T}}_{p,p}(\lambda)$  and
after some cumbersome simplifications we find that the final
result for $\Lambda(\lambda,\overrightarrow{\mu})$ is,
\begin{eqnarray}
\Lambda(\lambda,\overrightarrow{\mu})& = &\prod_{i=1}^{N}
S(\lambda,\mu_{i})_{1,1}^{1,1} \prod_{l=1}^{m} \frac{S(\lambda_l,
\lambda)_{1,1}^{1,1}}{S(\lambda_l,\lambda)_{2,1}^{2,1}} +
\sum_{k=2}^{n_{s}-1} \prod_{i=1}^{N}
S(\lambda,\mu_{i})_{k,1}^{k,1} \prod_{l=1}^{m}
{P_{k}(\lambda,\lambda_{l})} 
\nonumber \\
&& +\prod_{i=1}^{N}
S(\lambda,\mu_{i})_{n_s,1}^{n_s,1} \prod_{l=1}^{m}
\frac{S(\lambda,\lambda_l)^{n_{s},2}_{n_{s},2}}{S(\lambda,\lambda_l)_{n_{s},1}^{n_{s},1}},
\label{p12}
\end{eqnarray}
where
\begin{eqnarray}
P_{k}(\lambda,\mu)= \frac{ S(\lambda,\mu)_{k,2}^{k,2}
S(\lambda,\mu)_{k+1,1}^{k+1,1} - S(\lambda,\mu)_{k,2}^{k+1,1}
S(\lambda,\mu)_{k+1,1}^{k,2}} {S(\lambda,\mu)_{k,1}^{k,1}
S(\lambda,\mu)_{k+1,1}^{k+1,1}}
\end{eqnarray}

The remaining terms that are not proportional to the eigenvector
(\ref{state}) can be canceled out by imposing further restriction
on the rapidities $\{ \lambda_j \}$.  These are auxiliary Bethe
ansatz equations and in our case only a unique relation is needed
to eliminated all such unwanted terms. It is given by,
\begin{equation}
\prod_{l=1}^{N}
\frac{S(\lambda_{j},\mu_{l})_{1,1}^{1,1}}{S(\lambda_{j},\mu_{l})_{2,1}^{2,1}}=
\prod_{\stackrel{i=1}{i\neq j}}^{m}
\Theta(\lambda_{j},\lambda_{i})
\frac{S(\lambda_{j},\lambda_{i})_{1,1}^{1,1}}{S(\lambda_{j},\lambda_{i})_{2,1}^{2,1}}
\frac{S(\lambda_{i},\lambda_{j})_{2,1}^{2,1}}{S(\lambda_{i},\lambda_{j})_{1,1}^{1,1}}.
\label{bet}
\end{equation}

Before  proceeding we remark that our expressions (\ref{p12}-\ref{bet}) reproduce the eigenvalues
and Bethe ansatz equations of a particular integrable model with higher spin obtained through
fusion procedure of six-vertex $S$-matrices \cite{BAB}. Even in this special case, one hopes
that our general expression for the eigenstates (\ref{p11}) could still be of utility as
far as correlation functions are concerned \cite{MAI}.

At this point we have been able to derive that basic ingredients to
obtain the quantization rule for the one-particle momenta
$p(\mu_{j})$. This follows directly from
(\ref{BAE},\ref{p12}) and the unitarity
property of the $S$-matrix (\ref{p3}), i.e. the condition
$S(\lambda,\lambda)_{a,b}^{c,d} \sim \delta_{a,d} \delta_{b,c}$. The
final result is,
\begin{equation}
\mbox{e}^{ i p(\mu_{j}) L }=\prod_{\stackrel{i=1}{i\neq j}}^{N}
S(\mu_{j},\mu_{i})_{1,1}^{1,1} \prod_{l=1}^{m}
\frac{S(\lambda_l,\mu_j)_{1,1}^{1,1}}
{S(\lambda_l,\mu_j)_{2,1}^{2,1}}
\label{bet1}
\end{equation}

The results (\ref{bet},\ref{bet1}) are essential in order to investigate the
thermodynamic limit properties for a fixed density $N/L$. They have the merit
of being derived under mild assumptions for the two-body collision amplitudes
and therefore with an enormous potential of applicability.

In conclusion, we have solved exactly a system of spin-s particles
that interact via arbitrary $U(1)$ factorizable $S$-matrix. The
structure of both eigenvectors and the eigenvalues was derived
solely
from the Yang-Baxter relations (\ref{p2},\ref{p6}) and the
unitarity property (\ref{p3}). We
believe that our formula (\ref{p11}) is capable to accommodate the
solution of other integrable models possessing extra $U(1)$
symmetries other than that already discussed. In these cases, previous 
experience \cite{MA1}
suggests that the recursive structure of (\ref{p11})
will be preserved but now with
$\psi_{\frac{k}{2}}(\lambda_1,\lambda_{j_2},\dots,\lambda_{j_k})$
behaving as a vector-function 
while $\Theta(\lambda,\mu)$ as an underlying
factorizable $S$-matrix. If this proposal turns out to be
feasible in the future one would be able to solve exactly 
several distinct families of integrable models
from a rather unified point of view.

\section*{Acknowledgements}
The authors thank the Brazilian Research Agencies FAPESP and CNPq
for financial support.

\end{document}